\documentclass{iau} 
\usepackage{graphicx}

\title[Symposium 315~~The Smith Cloud] %% give here short title %%
{Accretion Onto the Milky Way: The Smith Cloud}

\author[Felix J. Lockman] {Felix J. Lockman}
 
\affiliation{National Radio Astronomy Observatory \footnote{The National Radio Astronomy Observatory is a facility of the National Science Foundation, operated under  cooperative agreement by
    Associated Universities, Inc.} \\ P.O. Box 2 \\
 Green Bank, WV 24944 USA \\ email: {\tt jlockman@nrao.edu} }

\pubyear{2015}
\volume{315}  %% insert here IAU Symposium No.
\jname{From interstellar clouds to star-forming galaxies: universal processes?}
\editors{}
\begin{document}

\maketitle

\begin{abstract}
Active gas accretion onto the Milky Way is observed in an object called
the Smith Cloud, which contains several million solar masses of neutral
and warm ionized gas and is currently losing material to the Milky Way,
 adding angular momentum to the disk.  
It is several kpc in size and its tip lies  two kpc
below the Galactic plane.  It appears to have 
no stellar counterpart, but could contain a stellar population 
like that of the dwarf galaxy Leo P.
There are suggestions that its existence
and survival require that it be embedded in a dark matter halo of a few
$10^8$ solar masses.

%% add here a maximum of 10 keywords, to be taken form the file <Keywords.txt>
\end{abstract}

\firstsection % if your document starts with a section,
              % remove some space above using this command.
\section{Introduction}

The puzzle of the origin of high velocity clouds (HVCs) is still with us more than 50 years after their initial 
discovery (\cite{Muller1963}),  and 
despite the enormous progress made in the past decades in understanding their distances, size, mass, 
metalicity, covering fraction, and ionization state (e.g., \cite{Wakker1997,Lockman2002,
Fox2006,Wakker2008,Thom2008,Shull2009,Putman2012,Lehner2012}).  
HVCs are now known to exist around nearby galaxies in a population that exends at least 50 kpc from the host system 
(\cite{Thilker2004,Grossi2008,Westmeier2008,Putman2009}.) 
The possibility that
{\it some}  HVCs may be intra-group or even intergalactic, or associated with dark matter halos, 
or tracers of  faint dwarf galaxies, cannot be excluded (e.g., \cite{Blitz1999,Braun1999,Adams2013}). 
Except for the Magellanic Stream,  the origin of even the most well-studied HVC is completely unknown.

\section{The Smith Cloud}

Perhaps the most exceptional HVC is the Smith Cloud (\cite{Smith1963}), 
first identified in observations at $35'$ 
resolution where it shows some head-tail structure, as do about 10\% of Milky Way HVCs 
(\cite{Bruns2000}).  Observations with the 100 meter Green Bank Telescope (GBT) at $9'$ resolution 
reveal a spectacular cometary cloud, showing extensive evidence of interaction between the cloud and 
the gaseous halo of the Milky Way (\cite{Lockman2008}).  Figure 1 shows an HI image  from new  unpublished 
GBT observations (Lockman et al. 2016).  The new data reveal some previously unknown cloud components 
symetrically placed  aside the main body of the cloud, as well as a very extended ``tail''. 

Much is now known about the Smith Cloud.  Its distance  
has been estimated from three independent techniques, 
which agree quite well (\cite{Lockman2008}).  Measurement of interstellar absorption lines 
against stars at varying distance provide the strongest constraint (\cite{Wakker2008}).
Basic properties of the cloud are listed in Table 1.  

Faint H$\alpha$ emission detected from the cloud suggests that it contains 
as much ionized as neutral gas (\cite{Bland-Hawthorn1998,Putman2003,Hill2009}).  
From a study of the rotation measure of background 
AGN \cite{Hill2013} deduce the presence of a magnetic field with an amplitude 
8 $\mu$G in the decelerated parts of the cloud.  New data, currently being analyzed, should improve 
both of these estimates.  

There is no obvious stellar counterpart to the cloud.  \cite{Stark2015} looked for young stars but did not 
find a statistically significant excess toward the cloud compared to adjacent areas.  
Contamination by the Milky Way at the low latitude 
of the cloud makes this work difficult, however, 
and Stark et al. emphasize that they cannot rule out a stellar population like that of Leo P, a  dwarf 
galaxy with a similar HI mass and  $M_V = -9.4$       
(\cite{McQuinn2013}).

\begin{figure}[b]
\vspace*{0.0 cm}
\begin{center}
\includegraphics[width=5in]{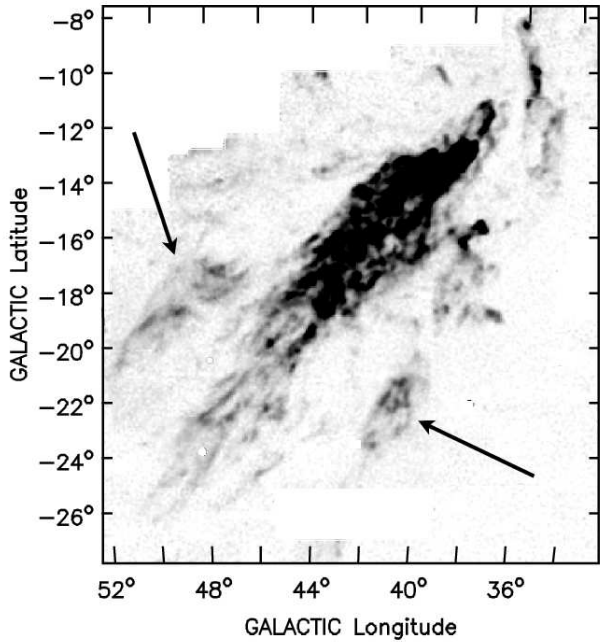}
%Fig1_SC2011_VGSR248_tr.pdf} 
%\vspace*{-0.0 cm}
 \caption{The Smith Cloud in HI from new unpublished observations with the GBT. 
The two features marked with arrows have the appearance of having been shed by the cloud as it moves 
toward the Galactic plane; each contains $\sim 5\times 10^4$ M$_{\odot}$. }
   \label{fig1}
\end{center}
\end{figure}

\begin{table}
  \begin{center}
  \caption{Smith Cloud Properties}
  \label{tab2}
 {\scriptsize
  \begin{tabular}{ll}\hline 
Distance$^{1,2}$ & $12.4 \pm 1.3$ kpc \\ 
R$_{gal}$ & $7.6 \pm 1.0$ kpc\\
z & $-2.2$ kpc \\
size & $\approx3 \times 1$ kpc \\
M$_{\rm HI} \ ^2$ & $2.0 \times 10^6$ M$_{\odot}$ \\
M$_{\rm H+} \ ^3$ & $>1.0 \times 10^6$ M$_{\odot}$ \\
B$_{||} \ ^4$ & $8\  \mu$G \\
 \end{tabular}
  }
 \end{center}
\vspace{1mm}
 \scriptsize{
 {\it Notes:} 
  1) Wakker et al. 2008.
 2) Lockman et al. 2008, 2016.
  3) Hill et al. 2009.  
  4) Hill et al. 2013.
}
\end{table}

\section{Interaction}

Evidence of interaction with the Milky Way abounds.  The overall morphology of the Smith Cloud 
suggests that it is moving towards the Galactic plane and enountering the Galactic halo, shedding 
pieces of itself as it goes. 
The tip of the cloud at $z \approx -2$ kpc is closer to the Galactic plane than the top of some 
HI superbubbles (\cite{Pidopryhora2007}). The densest portions of the cloud lie closest to the Galactic plane 
while the ``tail'', $10^{\circ}$ further from the  plane,  is much more diffuse.   There is kinematic 
evidence  of the interaction as well. 
 Figure 2 shows a velocity-position cut across the center of 
the cloud.  The main body of the cloud is distinct from Galactic HI, but there 
are kinematic ``bridges'' between the cloud edges and lower-velocity disk/halo gas.  
Here we see direct evidence that portions of the cloud have been decelerated and are now 
blending with the Galaxy.  There is also a ridge of decelerated gas conforming to the upper edge of the 
cloud (see Figure 4 of Lockman et al. (2008)).  
Several holes in the main body of the cloud at $V_{LSR} \approx 100$ km s$^{-1}$ correspond to 
$\sim 100\  M_{\odot}$  HI clumps at $\sim50$ km s$^{-1}$ 
lower $V_{LSR}$, suggesting that 
pieces of the cloud have broken away and been decelerated by the medium it 
is encountering.   There seems to be no doubt that 
the Smith Cloud is being destroyed by its interaction with the Milky Way and is losing gas to the 
Galaxy. A similar pattern of accretion is observed in the HVC Complex H (\cite{Lockman2003}).

\begin{figure}[b]
\vspace{-0.1 cm}
\begin{center}
\includegraphics[width=3.5in]{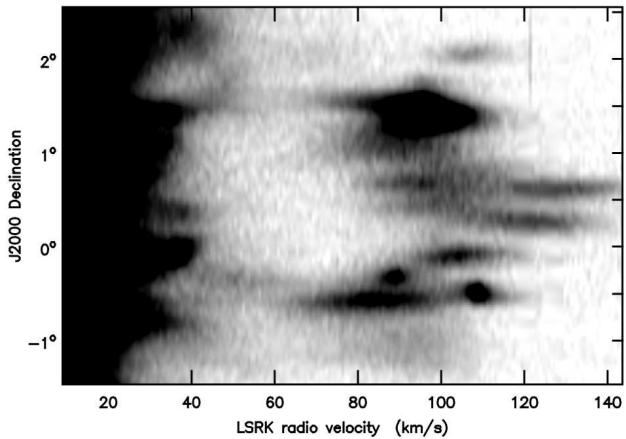}
\vspace*{-0.1 cm}
 \caption{Position-velocity cut across the minor axis of the Smith Cloud.  
The main part of the cloud lies at $V_{LSR} \approx 100$ km s$^{-1}$, but its edges 
show kinematic ``bridges'' to the lower-velocity Galactic gas, indicating that portions of the cloud 
are decelerating and being accreted by the Galaxy. }
   \label{fig2}
\end{center}
\end{figure}

\section{Trajectory and Progenitor}

\cite{Lockman2008} used the change of $V_{LSR}$ along the cloud's major axis to estimate a trajectory and 
found that the cloud has a total space velocity $\approx 300$ km s$^{-1}$, well below the escape velocity of the 
Milky Way.  While this analysis was based on early data and does not include the full extent of the cloud 
 in Fig.~1, there seems little doubt that the cloud's largest velocity component 
is  in the direction of Galactic rotation -- as it interacts with the Milky Way it therefore 
 adds angular momentum to the disk. 

It it likely that the main body of the cloud   was never more than 4 kpc from the Galactic plane
and its ``tail'' is  considerably closer to us than its head, 
though its exact orientation with respect to the plane of the sky is unknown.  There is 
even evidence for a extended component of the cloud ahead of the main body 
 (\cite{Lockman2012}).  Clarifying these points is clearly important to understanding the origin and 
fate of the cloud.  

Simulations suggest that clouds falling into the Milky Way halo are easily disrupted 
(\cite{Putman2012} and references therein), so 
given its striking appearance, evidence for interaction, and rather modest total 
space velocity,  a natural question is why the Smith Cloud has survived as long as it has.  
 One answer, 
first proposed by \cite{Nichols2009} and presented in more detail by \cite{Nichols2014} 
is that the Smith Cloud is the baryonic component of a dark matter sub-halo.
 A dark matter halo $\sim 3\times 10^8$ M$_{\odot}$ is 
sufficient to stabilize the cloud core while not massive enough to induce significant star 
formation, which is not observed.  This model may be entirely consistent with the Smith Cloud being a  
dwarf galaxy similar to Leo P, but a dwarf  whose accretion and disruption by the Milky Way are now being witnessed  
at close range.  

The Smith Cloud is an extraordinary object whose study may illuminate many aspects of the growth 
and evolution of galaxies.

\end{document}